\documentclass{article}
\pdfoutput=1
\usepackage[utf8]{inputenc}
\usepackage[dvips]{graphicx}
\usepackage{amsmath,amsfonts,amssymb,amsthm,physics}
\usepackage[a4paper, total={6in, 8in}]{geometry}
\usepackage{authblk}
\usepackage{hyperref}
\usepackage{url}
\usepackage[
backend=biber,
style=ieee,
sorting=none
]{biblatex}
\title{Crosscap states in the XXX spin-1/2 spin chain}
\author{Christopher Ekman\\cgekman@kth.se}
\affil{Department of Physics, KTH Royal Institute of Technology, SE-106 91 Stockholm, Sweden}
\date{July 2022}

\begin{document}

\maketitle
\begin{abstract}
    We consider integrable boundary states in the XXX spin-1/2 spin chain. We begin by briefly reviewing the algebraic Bethe Ansatz as well as integrable boundary states in spin chains. Then a recently discovered class of integrable states known as crosscap states is described and expanded. In these states each spin is entangled with its antipodal spin. We present a novel proof of the integrability of both a crosscap state that is known in the literature and one that has not previously been studied. We then use the machinery of the algebraic Bethe Ansatz to derive the overlaps between the crosscap states and off-shell Bethe states.
\end{abstract}
\section{Introduction}
In the last decade there has been an interest in obtaining exact formulae for overlaps between particular spin-chain states and eigenstates of integrable spin chains. The reason for this is twofold. First, a lot of progress has recently been made in the study of quantum quenches (see e.g. \cite{Calabrese_2006,Gogolin_2016,Eisert_2015} and references therein). The idea is to take a macroscopic system in the ground state $\ket{\Psi_0}$ of some Hamiltonian $H_0$ and at $t=0$ change it to some other Hamiltonian $H$ and thereafter allow the system to time evolve unitarily. If $H$ has energy eigenvectors $\ket{n}$ with energies $E_n$ the time evolution of the expectation value of some observable is given by
\begin{equation}
    \langle\mathcal{O}\rangle(t)=\sum_{m,n}\bra{\Psi_0}\ket{m}\bra{m}\mathcal{O}\ket{n}\bra{n}\ket{\Psi_0}e^{i(E_n-E_m)t}.
\end{equation}
Clearly the overlaps $\bra{\Psi_0}\ket{m}$ are of some importance in the time evolution. In 2014 it was found that for integrable spin chains the overlaps can sometimes be concisely written as determinants \cite{Pozsgay_2014,Brockmann_2014}. Because of the simple determinant formula it was possible to take the thermodynamic limit, so it could be used as the initial state in a quantum quench, as was done in \cite{Wouters_2014}. This sparked an interest in the computation of overlaps for integrable spin chains in the condensed matter community, and it has since been shown that a number of overlaps can be written as determinants \cite{Jiang2020,gombor_2021,Gombor_2021/03,Foda_2016, Piroli_2017}. More overlaps still have been conjectured and then checked for small spin chains \cite{Pozsgay_2018,de-Leeuw_2015,Buhl_Mortensen_2016,caetano_2022,Kristjansen_2020}. The initial states for which the overlaps can be written as determinants are known as \textit{integrable states}.

Second, it was found that spin chain overlaps arise in $3+1$ dimensional gauge theories \cite{de-Leeuw_2015}. The connection between integrable spin chains and gauge theory was first discovered for $\mathcal{N}=4\ \mathrm{SYM}$, a highly symmetric relative of QCD, when the one-loop anomalous dimensions of traces of products of scalars was found to correspond to the eigenvalues of the Heisenberg model \cite{Minahan_2003}. More generally, the trace of some combination of scalars corresponds to a particular state in a spin chain. If a defect, such as a domain wall, is added, then these operators develop non-zero vacuum expectation values. Because of the underlying spin chain structure these vacuum expectation values are proportional to overlaps between eigenstates of the spin chain and a class of states called matrix product states (see \cite{deLeeuw2020} for a review).

Due to the combined interest of the high-energy theory and condensed matter theory communities it is clear that it is important to find more states for which the overlaps can be written in a simple determinant form. The most well-studied classes of integrable states are the valence bond states and the matrix product states. In both cases the overlaps generally take the form
\begin{equation} 
    \frac{\bra{B}\ket{\{\lambda\}}}{\sqrt{\bra{\{\lambda\}}\ket{\{\lambda\}}}}=\prod_{k=1}F(\lambda_k)\sqrt{\frac{\det G^+}{\det G^-}},
\end{equation}
where $\bra{B}$ is the integrable state, $\ket{\{\lambda\}}$ is an on-shell Bethe state with Bethe roots $\{\lambda\},$ $F(\lambda)$ is a function of one variable, and the matrices $G^+,G^-$ are related to the Gaudin matrix. For valence bond states this formula has been proven for $\mathfrak{gl}(n)$ symmetric spin chains \cite{gombor_2021}.

Recently, a new class of integrable states was found \cite{caetano_2022}. These are known as crosscap states, and were proposed as analogues to the crosscap states that appear in conformal field theory. The purpose of this paper is to investigate this new class of states and in particular to investigate the overlaps with eigenstates of the Heisenberg model.

The paper is structured as follows: first, in section \ref{background} we establish our conventions by briefly describing the Algebraic Bethe Ansatz as well as integrable boundary states. Second, in section \ref{CC} two classes of integrable crosscap states for the Heisenberg model are described, and in section \ref{overlaps} some formulas for the overlaps between the crosscap states and off-shell Bethe states. Finally, in section \ref{conclusions} we present our conclusions and suggest some directions for future research.
\section{Background material} \label{background}
\subsection{The Algebraic Bethe Ansatz}
We begin by briefly reviewing the Algebraic Bethe Ansatz, mostly using the notation of \cite{Korepinbook}. For more extensive reivews, see for example \cite{Korepinbook,Slavnovnotes,Faddevnotes}. The physical model that we will consider is the XXX spin-$1/2$ spin chain with Hamiltonian
\begin{equation} \label{Hamiltonian}
    H=\sum_{n=1}^L 1-P_{n,n+1},
\end{equation}
which is also known as the Heisenberg model \cite{Heisenberg_1928}. Here $L$ is the number of sites, and $P_{n,n+1}$ exchanges the spins at site $n$ and site $n+1$. This Hamiltonian may be diagonalized by using the Algebraic Bethe Ansatz, where we derive the Hamiltonian from an object known as the Lax operator. Following \cite{Faddevnotes} we use the Lax operator 
\begin{equation}\label{Heis_L}
    L_{n,a}(\lambda,\nu)=(\lambda-\nu-\frac{i}{2})I_{n,a}+iP_{n,a}=\begin{bmatrix}
    \lambda-\nu+iS^z_n&iS^-_n\\
    iS^+_n&\lambda-\nu-iS^z_N
    \end{bmatrix},
\end{equation}
which acts on the tensor product of the quantum space corresponding to site $n$ and an auxiliary space, as indicated by the subscripts. The matrix elements are the usual $\mathrm{SU}(2)$ spin operators acting on the quantum space corresponding to site $n$. In addition to the rapidity $\lambda$ we have also introduced an inhomogeneity $\nu$. Another important object is the $R$-matrix which we take to be
\begin{equation}\label{R_XXX}
    R(\lambda)_{a_1,a_2}=I_{a_1,a_2}+\frac{i}{\lambda}P_{a_1,a_2}
\end{equation}
which acts on two auxiliary spaces. Together the Lax operator and the $R$-matrix satisfy the $RLL$ equation
\begin{equation} \label{local_FCR}
    R_{a_1,a_2}(\lambda-\mu)L_{n,a_1}(\lambda)L_{n,a_2}(\mu)=L_{n,a_2}(\mu)L_{n,a_1}(\lambda)R_{a_1,a_2}(\lambda-\mu).
\end{equation}

The monodromy matrix is given by the product of all Lax operators:
\begin{equation}
    T_a(\lambda,\{\nu\})=L_{1,a}(\lambda-\nu_1)L_{2,a}(\lambda-\nu_2)...L_{L,a}(\lambda-\nu_L).
\end{equation}
Suppressing the inhomogeneities in the argument the monodromy matrix $T_a(\lambda)$ is defined as
\begin{equation} \label{mon_elements}
    T_a(\lambda)=\begin{bmatrix}A(\lambda)&B(\lambda)\\C(\lambda)&D(\lambda)\end{bmatrix},
\end{equation}
where each element acts on the tensor product of all the quantum spaces in the spin chain. It follows from the $RLL$ equation that the monodromy matrix satisfies the $RTT$ relation
\begin{equation} \label{FCR}
    R_{a_1,a_2}(\lambda-\mu)T_{a_1}(\lambda)T_{a_2}(\mu)=T_{a_2}(\mu)T_{a_1}(\lambda)R_{a_1,a_2}(\lambda-\mu),
\end{equation}
which implies the following set of commutation relations for the elements of the monodromy matrix:
\begin{equation} \label{yang_comm}
    [T^i_j(\lambda),T^k_l(\mu)]=g(\lambda,\mu)\left(T^k_j(\mu)T^i_l(\lambda)-T^k_j(\lambda)T^i_l(\mu)\right),
\end{equation}
where we introduced the notation $g(\lambda,\mu)=\frac{i}{\lambda-\mu}$.

Now, the auxiliary space is unphysical, so we can get rid of it by tracing over it. This leads us to defining the transfer matrix $\tau(\lambda)$:
\begin{equation}
    \tau(\lambda)=\Tr_a T_a(\lambda)=A(\lambda)+D(\lambda).
\end{equation}
One can show that the Hamiltonian \eqref{Hamiltonian} is contained in the transfer matrix, so diagonalizing the Hamiltonian amounts to diagonalizing the transfer matrix. This can be done by first defining a vacuum state $\ket{0}$ which we take to be the spin state with all spins pointing upwards. We say that a spin pointing downwards is excited. The vacuum state satisfies 
\begin{equation}
    A(\lambda)\ket{0}=a(\lambda)\ket{0}\quad D(\lambda)\ket{0}=d(\lambda)\ket{0}\quad C(\lambda)\ket{0}=0,
\end{equation}
where
\begin{equation} \label{Heis_vac_eigs}
    a(\lambda)=\prod_{k=1}^L(\lambda-\nu_k+\frac{i}{2}),\quad d(\lambda)=\prod_{k=1}^L(\lambda-\nu_k-\frac{i}{2})
\end{equation}
are the vacuum eigenvalues of the $A$ and $D$ operators.
This motivates us to interpret the $C$-operator as an annihilation operator. Next we define Bethe states as
\begin{equation}
    \ket{\{\lambda\}}=\prod_{n=1}^N B(\lambda_n)\ket{0},
\end{equation}
where each $B$-operator creates a magnon in the spin chain.
One can use the commutation relations implied by the $RTT$ equation to show that the state $\ket{\{\lambda\}}$ is an eigenstate of the transfer matrix if the rapidities $\{\lambda\}$ satisfy the Bethe equations:
\begin{equation}\label{Bethe_eqs}
    \frac{a(\lambda_n)}{d(\lambda_n)}=\prod_{k\neq n}\frac{f(\lambda_n,\lambda_k)}{f(\lambda_k,\lambda_n)}.
\end{equation}
where we have introduced the notation $f(\lambda,\mu)=1+\frac{i}{\lambda-\mu}$. If the rapidities satisfy the Bethe equations then the corresponding Bethe state is said to be on-shell.

In the discussion of crosscap states below the idea of splitting the spin chain in two will be important. The monodromy matrix is a product of all the Lax matrices over the length of the spin chain, but we can equally well perform the multiplication over two connected partitions of the spin chain separately. This leads to two monodromy matrices $T_1,T_2$ which are defined as
\begin{equation} \label{T_1_T_2}
    T_1(\lambda)=\prod_{k=1}^{L'} L_{n,1}(\lambda-\nu_n),\quad T_2(\lambda)=\prod_{k=L'+1}^L L_{n,1}(\lambda-\nu_n).
\end{equation}
Each partition of the spin chain will henceforth be referred to as spin chain 1 and spin chain 2, respectively. The monodromy matrix on the full spin chain is obtained by multiplying $T_1$ and $T_2$ so that
\begin{equation} 
    T(\lambda)=T_1(\lambda)T_2(\lambda).
\end{equation}
Each spin chain has its own vacuum and its own creation and annihilation operators, which are distinguished by the subscripts 1 and 2. Note that the elements of $T_1$ commute with the elements of $T_2$ since they operate on different Hilbert spaces The creation operator for the full spin chain can be obtained from the creation operators on the sub-chains by carrying out the matrix multiplication:
\begin{equation}
    B(\lambda)=A_1(\lambda)B_2(\lambda)+B_1(\lambda)D_2(\lambda).
\end{equation}
The Bethe states can be written in terms of Bethe states on each sub-chain 
\begin{equation}
    \prod_{n=1}^NB(\lambda_n)\ket{0}=\prod_{n=1}^N(A_1(\lambda_n)B_2(\lambda_n)+B_1(\lambda_n)D_2(\lambda_n))\ket{0}_1\ket{0}_2.
\end{equation}
When we expand the product each term will contain operators acting on either spin chain 1 or spin chain 2 such that only one operator corresponding to each rapidity $\lambda_k$ is present. For each term, we need to move all $A_1$:s through all $B_1$:s, and all $D_2$:s through all $B_2$:s. In order to represent the sum we obtain in a useful way we split the set of rapiditied $\{\lambda\}$ into two partitions $\{\lambda^{I}\}$ and $\{\lambda^{II}\}$ such that all rapidities in $\{\lambda^{I}\}$ appear in operators acting on spin chain 1, and all rapidities in $\{\lambda^{II}\}$ appear in operators acting on spin chain 2. Performing these maneuvers we obtain
\begin{align} \label{B_split}
    \prod_{n=1}^NB(\lambda_n)\ket{0}=\sum_{\{\lambda\}=\{\lambda^{I}\}\cup\{\lambda^{II}\}}&\prod_{j\in {I}}\prod_{k\in {II}}[a_2(\lambda_j^{I})d_1(\lambda_k^{II})f(\lambda_k^{II},\lambda_j^{I})]\nonumber\\&\cross\prod_{j\in {I}}B_1(\lambda^{I}_j)\ket{0}_1\prod_{k\in {II}}B_2(\lambda^{II}_k)\ket{0}_2.
\end{align}
Here the sum is over all partitions of $\{\lambda\}$ into $\{\lambda^{I}\}$ and $\{\lambda^{II}\}$ and the functions $a_2$ and $d_1$ are vacuum eigenvalues of the second and first part of the spin chain respectively. The most general version of this formula may be found in \cite{Korepinbook}.
\subsection{Integrable boundary states}
For simplicity we set all inhomogeneities to $0$ in this section.
Integrable boundary states are states that satisfy \cite{Piroli_2017}
\begin{equation}\label{integrable_state_definition}
    \left[\tau(\lambda)-\tau(-\lambda)\right]\ket{B}=0,
\end{equation}
where $\ket{B}$ is the integrable state under consideration. This implies a selection rule for its overlaps with on-shell Bethe states, namely the overlaps are non-zero only if the rapidities are parity invariant, so
\begin{equation} \label{paired_rapidities}
    \{\lambda\}=\{\lambda_1,\lambda_2,...,\lambda_{\frac{N}{2}},-\lambda_1,-\lambda_2,...,-\lambda_{\frac{N}{2}}\}.
\end{equation}
If one of the rapidities is $0$ then it will only appear once, and in that case there is an odd number of rapidities. The parity invariance implies that any integrable state $\ket{B}$ may have non-zero overlap only with on-shell Bethe states that have momentum $0$ or $\pi$. To see this, we write the total momentum of a Bethe state $\ket{\{\lambda\}}$ \cite{Faddevnotes}
\begin{equation} \label{total_momentum}
    P(\{\lambda\})=\frac{1}{i}\sum_{k=1}^N\ln \frac{\lambda_k+\frac{i}{2}}{\lambda_k-\frac{i}{2}}.
\end{equation}
Note that this expression is singular for $\lambda_k=\pm \frac{i}{2}$. But if both $\frac{i}{2}$ and $-\frac{i}{2}$ are present among the rapidities the singularities may be regularized, and one can show that the pair $\pm\frac{i}{2}$ contribute $\pi$ to the momentum \cite{kristjansen_2021}. Furthermore, the rapidity $\lambda_0=0$ also contributes $0,$ while all the other rapidities do not contribute to the momentum. Thus any integrable state may have non-zero overlap only with on-shell Bethe states that have momentum $0$ or $\pi$.

As we stated in the introduction the overlaps between integrable states and on-shell Bethe states generally take the form
\begin{equation} \label{gen_int_overlap}
    \frac{\bra{B}\ket{\{\lambda\}}}{\sqrt{\bra{\{\lambda\}}\ket{\{\lambda\}}}}=\prod_{k=1}F(\lambda_k)\sqrt{\frac{\det G^+}{\det G^-}}.
\end{equation}
For the XXX spin-$1/2$ spin chain the matrices $G^{\pm}$ are given by
\begin{align}
     G^{\pm}_{kl}=\delta_{kl}\left[i\frac{\partial}{\partial \lambda_k}\ln\frac{a(\lambda_l)}{d(\lambda_l)}+\sum_{n=1}^N K^{+}(\lambda_k,\lambda_n)\right]-K^{\pm}(\lambda_k,\lambda_l)
\end{align}
where 
\begin{equation}
    K^{\pm}(\lambda,\mu)=\frac{2}{(\lambda-\mu)^2+1}\pm\frac{2}{(\lambda+\mu)^2+1}.
\end{equation}
To complete this section we introduce two quantities that will serve as building blocks for the crosscap overlaps below. First we write the scalar products between two Bethe states with $N$ magnons in a spin chain of length $L$ as
\begin{equation} \label{scalar_product}
    S_N^L(\{\lambda^C\},\{\lambda^B\})=\bra{0}\prod_{k=1}^NC(\lambda^C_k)\prod_{l=1}^N B(\lambda^B_l)\ket{0}.
\end{equation}
In general off-shell scalar products are quite complicated objects (see \cite{Korepinbook} for an extensive discussion). 
Second, we define
\begin{equation} \label{Z_L}
    Z_L=\bra{0'}\prod_{k=1}^LB(\lambda_k)\ket{0}
\end{equation}
where $\ket{0'}$ is the state with all spins pointing downwards. One can show that \cite{Korepin_1982}
\begin{equation} \label{partition_expression}
    Z_L(\{\lambda\},\{\nu\})=\frac{\prod_{k=1}^{L}\prod_{l=1}^L(\lambda_k-\nu_l+\frac{i}{2})(\lambda_k-\nu_l-\frac{i}{2})}{\prod_{1\leq k <l\leq L}(\nu_k-\nu_l)(\lambda_l-\lambda_k)}\mathrm{det}\mathcal{M},
\end{equation}
where $\mathcal{M}$ is a matrix whose elements are given by
\begin{equation} \label{partition_matrix}
    \mathcal{M}_{kl}=\frac{i}{(\lambda_k-\nu_l+\frac{i}{2})(\lambda_k-\nu_l-\frac{i}{2})}.
\end{equation}
\section{Crosscap states}\label{CC}
\subsection{General properties}
The crosscap states are states where antipodal spins are entangled. An arbitrary crosscap state can be written as
\begin{equation}
    \ket{\Psi}=\prod_{n=1}^{\frac{L}{2}}\ket{\psi}_{n,n+\frac{L}{2}}^{\otimes},
\end{equation}
where the subscripts $n,n+\frac{L}{2}$ indicates which sites are contained in each $\ket{\psi}$. The pieces $\ket{\psi}$ can be parametrized in the following way:
\begin{equation}
    \ket{\psi}=(\cos\beta+\nu)\ket{\uparrow\downarrow}+(\cos\beta-\nu)\ket{\downarrow\uparrow}+ie^{i\alpha}\sin\beta\ket{\uparrow\uparrow}+e^{i\alpha}\sin\beta\ket{\uparrow\uparrow}.
\end{equation}
The crosscap state where $\nu=0,\beta=\frac{\pi}{2}$ was introduced and shown to be integrable in \cite{caetano_2022}. Explicitly this state is given by
\begin{equation}
    \ket{C}=\prod_{n=1}^{\frac{L}{2}}\ket{c}_{n}^{\otimes},
\end{equation}
where
\begin{equation}
    \ket{c}_{n}=\ket{\uparrow\uparrow}_{n,n+\frac{L}{2}}+\ket{\downarrow\downarrow}_{n,n+\frac{L}{2}}
\end{equation}
so antipodal spins are identified. In order to characterize this state we can note that it has momentum $0$. To see this one take an arbitrary spin chain state where antipodal spins are identified. If one operates on this state with the translation operator $U=e^{iP}$ we obtain a state where antipodal spins are still identified. Hence $U\ket{C}=\ket{C}$ so the momentum is $0$. This distinguishes $\ket{C}$ from valence bond states, since the valence bond states necessarily contain components with both momentum $0$ and momentum $\pi$.

Another integrable crosscap state that has not appeared in the literature previously is the crosscap singlet, which is obtained by taking the limit $\nu\to\infty$. Renormalizing by dividing by $\nu$ we obtain
\begin{equation}
    \ket{C_S}=\prod_{n=1}^{\frac{L}{2}}\ket{c_s}_n^{\otimes},
\end{equation}
where
\begin{equation}
    \ket{c_s}_n=(\ket{\uparrow\downarrow}_{n,n+\frac{L}{2}}-\ket{\downarrow\uparrow}_{n,n+\frac{L}{2}})
\end{equation}
The proof that it is integrable will be presented in section \ref{sect:cross_int}. The crosscap singlet consists of a set of $\mathrm{SU}(2)$ singlets where the entangled spins are antipodal. Exactly half of the spins are excited in $\ket{C_S}$, with the excited spins being spread out over each half of the spin chain. Each term in $\ket{C_S}$ has a sign $(-1)^{n_1}$ where $n_1$ is the number of excited spins on the first half of the spin chain. If we operate on the crosscap singlet with the translation operator the two sides of the spin chain exchange one excited spin, so
\begin{equation}
    U\ket{C_S}=-\ket{C_S}.
\end{equation}
We see that similarly to $\ket{C}$ the state $\ket{C_S}$ is a momentum eigenstate, with the difference that $\ket{C_S}$ has momentum $\pi$. This makes it uniquely distinguished among all integrable states: the valence bond states are not momentum eigenstates at all \cite{Piroli_2017}, and the matrix product states all have momentum $0$ (see e.g. \cite{de-Leeuw_2015}), so the crosscap singlet is the only known integrable boundary state that has momentum $\pi$. 

\subsection{Proof of integrability} \label{sect:cross_int}
In order to prove that the two crosscap states are integrable we use the Lax operator given by equation \eqref{Heis_L}. Then the following relations can be easily checked
\begin{equation} \label{cc_lax_relation}
    \sigma^2L_{n,a}(\lambda)\sigma^2\ket{c}_n = -L_{n+\frac{L}{2},a}(-\lambda)\ket{c}_n,\quad L_{n,a}(\lambda)\ket{c_s}_n = -L_{n+\frac{L}{2},a}(-\lambda)\ket{c_s}_n,
\end{equation}
where the Pauli matrix $\sigma^2$ acts on the auxiliary space. Suppose $T_1(\lambda,\{\nu\}_1)$ is the monodromy matrix acting on all the sites $1$ to $\frac{L}{2}$ with inhomogeneities $\{\nu\}_1$ and $T_2(\lambda,\{\nu\}_2)$ is the monodromy matrix acting on all the spins with $\frac{L}{2}+1$ to $L$ with inhomogeneities $\{\nu\}_2$. Then equations \eqref{cc_lax_relation} implies the equations
\begin{align} \label{cc_T_relation1}
    \sigma^2 T_1(\lambda,\{\nu\}_1)\sigma^2\ket{C}&=(-1)^{\frac{L}{2}}T_2(-\lambda,\{-\nu\}_1)\ket{C},\\\label{cc_T_relation2} T_1(\lambda)\ket{C_S}&=(-1)^{\frac{L}{2}}T_2(-\lambda,\{-\nu\}_1)\ket{C_S}.
\end{align}
Note that when $\lambda\to-\lambda$ all the inhomogeneities also change sign. We will suppress the inhomogeneity arguments below to improve readability.
We prove integrability for the state $\ket{C}$ using a different method than the one that was used in \cite{caetano_2022}. Our argument can easily be adapted to $\ket{C_S}$ by replacing $\sigma^2$ with the identity below. In order to prove that $\ket{C}$ is integrable we need to show that
\begin{equation}
   (\tau(\lambda)-\tau(-\lambda))\ket{C}=0\iff \Tr T(\lambda)\ket{C}=\Tr T(-\lambda)\ket{C}.
\end{equation}
The trace is over the auxiliary space. Recalling that $T(\lambda)=T_1(\lambda)T_2(\lambda)$ we can use equation \eqref{cc_T_relation1} to find
\begin{align}
    \Tr T(\lambda,\{\nu\})\ket{C}&=\Tr\left(T_1(\lambda,\{\nu\}_1)T_2(\lambda,\{\nu\}_2)\right)\ket{C}\nonumber\\&=(-1)^{\frac{L}{2}}\Tr\left(T_1(\lambda,\{\nu\}_1)\sigma^2 T_1(-\lambda,\{-\nu\}_2)\sigma^2 \right)\ket{C}.
\end{align}
In order to use the algebraic structure of the monodromy matrix we need to assume that $T_1(\lambda,\{\nu\}_1)$ and $T_1(-\lambda,\{-\nu\}_2)$ are operating on the same spin chain. This means that the inhomogeneities must satisfy \begin{equation}
    \nu_{n+\frac{L}{2}}=-\nu_n.
\end{equation}
Assuming that this is true, we write the product of monodromy matrices using index notation (with summation over repeated indices) and use equation \eqref{yang_comm} twice to get
\begin{align}
    \Tr[T_1(\lambda),\sigma^2 T_1(-\lambda)\sigma^2 ]&=(\sigma^2)^j_k (\sigma^2)^l_i[(T_1)^i_j(\lambda),(T_1)^k_l(-\lambda)]\nonumber\\
    &=g(\lambda,-\lambda)(\sigma^2)^j_k(\sigma^2)^l_i\nonumber\\ &\cross \left((T_1)^k_j(-\lambda)(T_1)^i_l(\lambda)-(T_1)^k_j(\lambda)(T_1)^i_l(-\lambda)\right)\nonumber\\
    &=g(\lambda,-\lambda)(\sigma^2)^j_k(\sigma^2)^l_i[(T_1)^k_j(-\lambda),(T_1)^i_l(\lambda)]\nonumber\\
    &=g(\lambda,-\lambda)g(-\lambda,\lambda)(\sigma^2)^j_k(\sigma^2)^l_i\nonumber\\ &\ \cross\left((T_1)^i_j(\lambda)(T_1)^k_l(-\lambda)-(T_1)^i_j(-\lambda)(T_1)^k_l(\lambda)\right)\nonumber\\
    &=g(\lambda,-\lambda)g(-\lambda,\lambda)(\sigma^2)^j_k(\sigma^2)^l_i[(T_1)^i_j(\lambda),(T_1)^k_l(-\lambda)],
\end{align}
where we renamed the indices in the third and last steps in order to produce the commutators. Hence 
\begin{equation}
    \Tr[T_1(\lambda),\sigma^2 T_1(-\lambda)\sigma^2 ]=g(\lambda,-\lambda)g(-\lambda,\lambda)\Tr[T_1(-\lambda),\sigma^2 T_1(\lambda)\sigma^2 ].
\end{equation}
Assuming that $g(\lambda,-\lambda)g(-\lambda,\lambda)\neq 1$ this implies that
\begin{equation}
    \Tr[T_1(\lambda),\sigma^2 T_1(-\lambda)\sigma^2 ]=0\iff \Tr\left(T_1(\lambda)\sigma^2 T_1(-\lambda)\sigma^2\right)=\Tr\left(T_1(-\lambda)\sigma^2 T_1(\lambda)\sigma^2\right)
\end{equation}
so finally using equation \eqref{cc_T_relation1} once more and restoring the inhomogeneities we obtain
\begin{align}
    (-1)^{\frac{L}{2}}\Tr\left(T_1(\lambda)\sigma^2 T_1(-\lambda)\sigma^2\right)\ket{C}&=(-1)^{\frac{L}{2}}\Tr\left(T_1(-\lambda)\sigma^2 T_1(\lambda)\sigma^2\right)\ket{C}\nonumber\\
    &=\Tr\left(T_1(-\lambda)\sigma^2 T_1(\lambda)\sigma^2\right)\ket{C}\nonumber\\
    &= \Tr\left(T_1(-\lambda)T_2(-\lambda)\right)\ket{C}\nonumber\\
    &=\Tr\left(T(-\lambda)\right)\ket{C}.
\end{align}
We have now shown that if the inhomogeneities are paired such than $\nu_{n+\frac{L}{2}}=-\nu_n$ then
\begin{equation}
    (\tau(\lambda)-\tau(-\lambda))\ket{C}=0
\end{equation}
whenever $g(\lambda,-\lambda)g(-\lambda,\lambda)\neq 1$. Recalling that $g(\lambda,\mu)=\frac{i}{\lambda-\mu}$ we see that this is equivalent to the requirement $\lambda\neq \pm\frac{1}{2}$. This is not a commonly occuring value for rapidities, so it is curious that the proof does not work for in this case. The proof in \cite{caetano_2022} does work for $\lambda= \pm\frac{1}{2}$, but not for $\lambda= \pm i$ (see their equation (4.29) and the ones below), so our proof complements theirs.

Note that changing the $\sigma^2$ to anything other than the identity either leads to non-integrable states or integrable states that are global rotations of $\ket{C}$. The states $\ket{C}$ and $\ket{C_S}$ therefore represent the only two classes of integrable crosscap states in the XXX spin-1/2 spin chain.
\section{Overlap formulae} \label{overlaps}
\subsection{On-shell overlaps} \label{sect:on-shell_overlaps}
In \cite{caetano_2022} it was conjectured on the basis of integrability and the general form \eqref{gen_int_overlap} that the crosscap state $\ket{C}$ has the overlap
\begin{equation}
    \frac{\bra{C}\ket{\{\lambda\}}}{\sqrt{\bra{\{\lambda\}}\ket{\{\lambda\}}}}=\sqrt{\frac{\det G^+}{\det G^-}}.
\end{equation}
This formula has been verified for small spin chains and small numbers of magnons. 

For the crosscap singlet state $\ket{C_S}$ no successful guess based on the form \eqref{gen_int_overlap} has been made. As was stated in the previous section $\ket{C_S}$ is unique among the integrable boundary states by having momentum $\pi$. On the basis of the discussion below equation \eqref{paired_rapidities} this means that $\ket{C_S}$ has non-zero overlap only with parity invariant on-shell Bethe states that also have either rapidities $\pm\frac{i}{2}$ or a rapidity equal to $0$. In the former case the Bethe state is singular and requires special care \cite{Heisenberg_1928,Nepomechie_2013,Rahul1998,avdeev}. For example, one can show that for singular Bethe states the rapidities must satisfy the following equation \cite{Nepomechie_2013}
\begin{equation} \label{sing_condition}
    \left[-\prod_{k=3}^N\frac{\lambda_k+\frac{i}{2}}{\lambda_k-\frac{i}{2}}\right]^N=1
\end{equation}
in addition to the Bethe equations. This severely restricts the Bethe roots. Finally exactly half the spins in $\ket{C_S}$ are excited, so if it is to have non-zero overlap with a Bethe state then that Bethe state must have $\frac{L}{2}$ magnons.
\subsection{Off-shell formulae for the crosscap overlaps}
We now move on to discuss some novel results regarding the crosscap overlaps. 

We wish to calculate quantities of the form 
\begin{equation}
    \bra{C}\prod_{k=1}^{N}B(\lambda_k)\ket{0}.
\end{equation}
Recall equations \eqref{cc_T_relation1} and \eqref{cc_T_relation2}.These relations suggest that it is productive to split the spin chain into two halves. Using equation \eqref{B_split} we get
\begin{align} \label{cc_overlap}
    \bra{C}\prod_{n=1}^NB(\lambda_n)\ket{0}=\bra{C}\sum_{\{\lambda\}=\{\lambda^{I}\}\cup\{\lambda^{II}\}}&\prod_{j\in {I}}\prod_{k\in {II}}[a_2(\lambda_j^{I})d_1(\lambda_k^{II})f(\lambda_k^{II},\lambda_j^{I})]\nonumber\\&\cross\prod_{j\in {I}}B_1(\lambda^{I}_j)\ket{0}_1\prod_{k\in {II}}B_2(\lambda^{II}_k)\ket{0}_2
\end{align}
where the sum is over all partitions of the set of rapidities into two sets.
From equation \eqref{cc_T_relation1} we see that 
\begin{equation}
    \bra{C}B_1(\lambda)=(-1)^{\frac{L}{2}+1}\bra{C}C_2(-\lambda).
\end{equation}
This permits us to turn all $B_1$ operators into $C_2$ operators under the assumption that the inhomogeneities satisfy $\nu_{n+\frac{L}{2}}=-\nu_n$ as in the preceding section. Hence
\begin{align}
    \bra{C}\prod_{n=1}^NB(\lambda_n)\ket{0}=\bra{C}\sum_{\{\lambda\}=\{\lambda^{I}\}\cup\{\lambda^{II}\}}&\prod_{j\in {I}}\prod_{k\in {II}}[a_2(\lambda_j^{I})d_1(\lambda_k^{II})f(\lambda_k^{II},\lambda_j^{I})]\nonumber\\&\cross(-1)^{n_{I}(\frac{L}{2}+1)}\ket{0}_1\prod_{j\in {I}}C_2(-\lambda^{I}_j)\prod_{k\in {II}}B_2(\lambda^{II}_k)\ket{0}_2,
\end{align}
where $n_{I}$ is the cardinality of the set $\{\lambda^I\}$.
The only term in $\bra{C}$ that contributes is the one where none of the spins on the first half are excited. From the definition of $\ket{C}$ we see that this term is $\bra{0}=\bra{0}_1\bra{0}_2$, so
\begin{align}
    \bra{C}\prod_{n=1}^NB(\lambda_n)\ket{0}=\sum_{\{\lambda\}=\{\lambda^{I}\}\cup\{\lambda^{II}\}}&\prod_{j\in {I}}\prod_{k\in {II}}[a_2(\lambda_j^{I})d_1(\lambda_k^{II})f(\lambda_k^{II},\lambda_j^{I})]\nonumber\\&\cross(-1)^{n_{I}(\frac{L}{2}+1)}\bra{0}_2\prod_{j\in {I}}C_2(-\lambda^{I}_j)\prod_{k\in {II}}B_2(\lambda^{II}_k)\ket{0}_2.
\end{align}
The factor involving creation and annihilation operators is the scalar product between two off-shell Bethe states:
\begin{equation}
    \bra{0}_2\prod_{j\in {I}}C_2(-\lambda^{I}_j)\prod_{k\in {II}}B_2(\lambda^{II}_k)\ket{0}_2=S_{N}^{\frac{L}{2}}(\{-\lambda^I\},\{\lambda^{II}\})
\end{equation}
where the superscript indicates that the scalar product is only on one half of the spin chain. Hence
\begin{align} 
    \bra{C}\prod_{n=1}^NB(\lambda_n)\ket{0}=\sum_{\{\lambda\}=\{\lambda^{I}\}\cup\{\lambda^{II}\}}&\prod_{j\in {I}}\prod_{k\in {II}}[a_2(\lambda_j^{I})d_1(\lambda_k^{II})f(\lambda_k^{II},\lambda_j^{I})]\nonumber\\&\cross(-1)^{n_{I}(\frac{L}{2}+1)}S_{N}^{\frac{L}{2}}(\{-\lambda^I\},\{\lambda^{II}\}).\label{cc_off-shell}
\end{align}
Note that if the two partitions $\{\lambda^I\}$ and $\{\lambda^{II}\}$ have different cardinalities the scalar product is $0$, so the sum is over all partitions that have equal cardinality $n_{I}=n_{II}=\frac{N}{2}$. Due to the complexity of off-shell scalar products this expressions has not been simplified further.

A similar calculation can be performed for the crosscap singlet state $\ket{C_S}$. From equation \eqref{cc_T_relation2} we have
\begin{equation}
    \bra{C_S}B_1(\lambda)=(-1)^{\frac{L}{2}}\bra{C_S}B_2(-\lambda)
\end{equation}
so
\begin{align}
    \bra{C_S}\prod_{n=1}^NB(\lambda_n)\ket{0}&=\bra{C_S}\sum_{\{\lambda\}=\{\lambda^{I}\}\cup\{\lambda^{II}\}}\prod_{j\in {I}}\prod_{k\in {II}}[a_2(\lambda_j^{I})d_1(\lambda_k^{II})f(\lambda_k^{II},\lambda_j^{I})]\nonumber\\&\cross\prod_{j\in {I}}B_1(\lambda^{I}_j)\ket{0}_1\prod_{k\in {II}}B_2(\lambda^{II}_k)\ket{0}_2\nonumber\\&=\bra{C_S}\sum_{\{\lambda\}=\{\lambda^{I}\}\cup\{\lambda^{II}\}}\prod_{j\in {I}}\prod_{k\in {II}}[a_2(\lambda_j^{I})d_1(\lambda_k^{II})f(\lambda_k^{II},\lambda_j^{I})]\nonumber\\&\cross(-1)^{\frac{n_IL}{2}}\ket{0}_1\prod_{j\in {I}}B_2(-\lambda^{I}_j)\prod_{k\in {II}}B_2(\lambda^{II}_k)\ket{0}_2.
\end{align}
The only contributing term from $\bra{C_S}$ is again the one where none of the spins on the first half of the spin chain are excited. From the definition of $\ket{C_S}$ we see that this term is $\bra{0}_1\bra{0'}_2$ where $\bra{0'}$ is the state where all the spins are excited. Hence
\begin{align}
    \bra{C_S}\prod_{n=1}^NB(\lambda_n)\ket{0}=\sum_{\{\lambda\}=\{\lambda^{I}\}\cup\{\lambda^{II}\}}&\prod_{j\in {I}}\prod_{k\in {II}}[a_2(\lambda_j^{I})d_1(\lambda_k^{II})f(\lambda_k^{II},\lambda_j^{I})]\nonumber\\&\cross(-1)^{\frac{n_IL}{2}}\bra{0'}_2\prod_{j\in {I}}B_2(-\lambda^{I}_j)\prod_{k\in {II}}B_2(\lambda^{II}_k)\ket{0}_2.
\end{align}
The factor involving annihilation operators was defined in equation \eqref{Z_L}:
\begin{equation}
    \bra{0'}_2\prod_{j\in {I}}B_2(-\lambda^{I}_j)\prod_{k\in {II}}B_2(\lambda^{II}_k)\ket{0}_2=Z_{\frac{L}{2}}(\{-\lambda^I\}\cup\{\lambda^{II}\},\{\nu\}_2).
\end{equation}
The overlap formula becomes
\begin{align} \label{singlet_off-shell}
    \bra{C_S}\prod_{n=1}^NB(\lambda_n)\ket{0}=\sum_{\{\lambda\}=\{\lambda^{I}\}\cup\{\lambda^{II}\}}&\prod_{j\in {I}}\prod_{k\in {II}}[a_2(\lambda_j^{I})d_1(\lambda_k^{II})f(\lambda_k^{II},\lambda_j^{I})]\nonumber\\&\cross (-1)^{\frac{n_IL}{2}}Z_{\frac{L}{2}}(\{-\lambda^I\}\cup\{\lambda^{II}\},\{\nu\}_2).
\end{align}
Note that $Z_{\frac{L}{2}}$ is $0$ if the number of magnons differs from $\frac{L}{2}$. The crosscap singlet overlap is therefore non-zero only if $N=\frac{L}{2}$.

In contrast to the crosscap state overlap as shown in formula \eqref{cc_off-shell} this formula can be simplified drastically. But first we will discuss some recursion properties of the crosscap singlet overlap To simplify the notation we define
\begin{equation}
    \bra{C_S}\prod_{n=1}^{\frac{L}{2}}B(\lambda_n)\ket{0}=O_L(\{\lambda\},\{\nu\})
\end{equation}
First, both $a_2$ and $d_1$ are symmetric in the inhomogeneities. Since $Z_L$ is symmetric in the inhomogeneities this means that $O_L$ is also symmetric in the inhomogeneities. Furthermore, the sum is over all partitions of the set of rapidities. This coupled with the fact that $Z_L$ is symmetric in the rapidities implies that $O_L$ is symmetric in the rapidities as well. 

Second, $O_L$ is a polynomial of order $L-1$ in all its arguments. To see why it has order $L-1$ we recall that $Z_{\frac{L}{2}}$ is a polynomial with order $\frac{L}{2}-1$ in all its arguments. Each vacuum eigenvalue in equation \eqref{singlet_off-shell} is of order $\frac{L}{2}$ in its argument, so $O_L$ is of order $\frac{L}{2}+\frac{L}{2}-1=L-1$ in each rapidity. Each inhomogeneity appears once in each vacuum eigenvalue factor in equation \eqref{singlet_off-shell} and there are $\frac{L}{2}$ such factors, so $O_L$ is of order $\frac{L}{2}+\frac{L}{2}-1=L-1$ in each inhomogeneity. In order to show that $O_L$ is a polynomial we need to show that its residue at each pole is $0$. We know that $Z_{\frac{L}{2}}$ is a polynomial, so only the factors $f(\lambda_k^{II},\lambda_j^{I})$ contribute with poles whenever $\lambda_k\to\lambda_l$, $l\neq k$. The singular part involving $\lambda_1$ and $\lambda_2$ is
\begin{align}
    \frac{i}{\lambda_2-\lambda_1}&a_2(\lambda_1)d_1(\lambda_2)Z_{\frac{L}{2}}({-\lambda_1}\cup\{\lambda_2\}\{-\lambda^I\}\cup\{\lambda^{II}\},\{\nu\}_2)\nonumber\\&+ \frac{i}{\lambda_1-\lambda_2}a_2(\lambda_2)d_1(\lambda_1)Z_{\frac{L}{2}}({-\lambda_2}\cup\{\lambda_1\}\{-\lambda^I\}\cup\{\lambda^{II}\},\{\nu\}_2).
\end{align}
In the limit where $\lambda_1\to\lambda_2$ these two terms cancel, so the residue at this pole is $0$. Since all other poles have the same form this means that $O_L$ is a polynomial.

Third, we can obtain a recursion relation for $O_L$ by setting $\lambda_1=\nu_1+\frac{i}{2}$. This sets $d_1(\lambda_1)=0$, as can be seen in equation \eqref{Heis_vac_eigs}. The rapidity $\lambda_1$ is therefore forced to be in partition $I$ so
\begin{align}
    O_L\bigg\rvert_{\lambda_1=\nu_1+\frac{i}{2}}=a_2(\lambda_1)\sum_{\{\lambda\}=\{\lambda^{I}\}\cup\{\lambda^{II}\}}&\prod_{j\in {I}}\prod_{k\in {II}}[a_2(\lambda_j^{I})d_1(\lambda_k^{II})f(\lambda_k^{II},\lambda_1)f(\lambda_k^{II},\lambda_j^{I})]\nonumber\\&\cross (-1)^{\frac{(n_I+1)L}{2}}Z_{\frac{L}{2}}(\{-\lambda_1\}\cup\{-\lambda^I\}\cup\{\lambda^{II}\},\{\nu\}_2)
\end{align}
Identifying the vacuum eigenvalues in equation \eqref{partition_expression} we can write
\begin{equation}
    Z_L=\frac{\prod_{k=1}^La(\lambda_k)d(\lambda_k)}{\prod_{1\leq k <l\leq L}(\nu_k-\nu_l)(\lambda_l-\lambda_k)}\det\mathcal{M}.
\end{equation}
Furthermore using the relations $a_2(-\lambda)=(-1)^{\frac{L}{2}}d_1(\lambda),\ d_2(-\lambda)=(-1)^{\frac{L}{2}}a_1(\lambda)$
we then obtain the following recursion relation for $O_L$
\begin{equation} \label{singlet_recursion1}
    O_L\bigg\rvert_{\lambda_1=\nu_1+\frac{i}{2}}=a_2(\lambda_1)d_1(\lambda_1)O_{L-1}(\{\lambda\}/\lambda_1,\{\nu\}/\nu_1)^{\mathrm{mod}_1}
\end{equation}
where the superscript $^{\mathrm{mod}_1}$ indicates that the vacuum eigenvalues have been modified in the following way
\begin{align}
    a_1(\lambda)&\to a_1(\lambda)(\lambda-\nu_1+\frac{i}{2})\nonumber\\ d_1(\lambda)&\to d_1(\lambda)(\lambda+\nu_1-\frac{i}{2})f(\lambda,\lambda_1)\nonumber\\
    a_2(\lambda)&\to a_2(\lambda)\nonumber\\d_2(\lambda)&\to d_2(\lambda)\nonumber
\end{align}
One can obtain a similar recursion relation by setting $\lambda_1=-\nu_1-\frac{i}{2}$:
\begin{equation} \label{singlet_recursion2}
    O_L\bigg\rvert_{\lambda_1=-\nu_1-\frac{i}{2}}=d_2(\lambda_1)a_1(\lambda_1)O_{L-1}(\{\lambda\}/\lambda_1,\{\nu\}/\nu_1)^{\mathrm{mod}_2}
\end{equation}
where $^{\mathrm{mod}_2}$ indicates the following modification
\begin{align}
    a_1(\lambda)&\to a_1(\lambda)(\lambda-\nu_1+\frac{i}{2})f(\lambda_1,\lambda)\nonumber\\ d_1(\lambda)&\to d_1(\lambda)(\lambda+\nu_1-\frac{i}{2})\nonumber\\
    a_2(\lambda)&\to a_2(\lambda)\nonumber\\d_2(\lambda)&\to d_2(\lambda)\nonumber
\end{align}
These recursion relations give us $O_L$ at $L$ different values for each of its arguments, thus determining it completely if $O_2$ is known. One can obtain the initial value $O_2$ by direct calculation:
\begin{equation}
    O_2(\lambda,\{\nu,-\nu\})=\frac{d_1(\lambda)a_2(\lambda)d_2(\lambda)}{(\lambda+\nu-\frac{i}{2})(\lambda+\nu+\frac{i}{2})}-\frac{a_2(\lambda)a_1(\lambda)d_1(\lambda)}{(-\lambda+\nu-\frac{i}{2})(-\lambda+\nu+\frac{i}{2})}
\end{equation}
where we used $a_2(-\lambda)d_2(-\lambda)=a_1(\lambda)d_1(\lambda)$. No concise solution to this recuresion has been found. 
\subsection{A determinant formula for the singlet overlap}
The singlet overlap $O_L$ can be written in a concise form using auxiliary quantum fields. We begin by carrying out a multi-row Laplace expansion on the rows corresponding to $\{-\lambda^{I}\}$ in the determinant $\det \mathcal{M}$ in $Z_{\frac{L}{2}}((\{-\lambda_1\}\cup\{-\lambda^I\}\cup\{\lambda^{II}\},\{\nu\}_2))$. The columns correspond to the inhomogeneities, so we obtain
\begin{equation}
    \det\mathcal{M}(\{-\lambda^{I}\}\cup\{\lambda^{II}\},\{\nu\})=\sum_{\{\nu\}=\{\nu^{\gamma}\}\cup\{\nu^{\delta}\}}(-1)^{P}\det\mathcal{M}(\{-\lambda^{I}\},\{\nu^{\gamma}\})\det\mathcal{M}(\{\lambda^{II}\},\{\nu^{\delta}\}),
\end{equation}
where the sum is over all partitions of the set of inhomogeneities such that the cardinality of $\{\nu^{\gamma}\}$ is equal to the cardinality of $\{\lambda^I\}$. The sign $(-1)^P$ is the sign of the permutation corresponding to the particular partition of $\{\lambda\},\{\nu\}$. Inserting this relation into expression \eqref{singlet_off-shell} and factoring out the prefactor in $Z_{\frac{L}{2}}$  we get
\begin{align} \label{singlet_det1}
    \bra{C_S}\prod_{n=1}^{\frac{L}{2}}B(\lambda_n)\ket{0}=&\frac{\prod_{k=1}^N a_2(\lambda_k)d_2(\lambda_k)}{\prod_{1\leq k <l\leq \frac{L}{2}}(\nu_l-\nu_k)(\lambda_l-\lambda_k)}\nonumber\\&\cross\sum_{\substack{\{\lambda\}=\{\lambda^{I}\}\cup\{\lambda^{II}\}\\ \{\nu\}=\{\nu^{\gamma}\}\cup\{\nu^{\delta}\}}}(-1)^P\prod_{j\in {I}}\prod_{k\in {II}}[a_2(\lambda_j^{I})d_1(\lambda_k^{II})f(\lambda_k^{II},\lambda_j^{I})]\nonumber\\&\cross (-1)^{\frac{n_IL}{2}}\prod_{j\in {I}}\prod_{k\in {II}}\left[\frac{\lambda_j^{I}-\lambda_{k}^{II}}{\lambda_j^{I}+\lambda_{k}^{II}}\right]\prod_{j\in I}\frac{a_1(\lambda^I_j)d_1(\lambda^I_j)}{a_2(\lambda^I_j)d_2(\lambda^I_j)}\nonumber\\&\cross \det\mathcal{M}(\{-\lambda^{I}\},\{\nu^{\gamma}\})\det\mathcal{M}(\{\lambda^{II}\},\{\nu^{\delta}\}).
\end{align}
This expression may be simplified by making use of the Laplace formula for the determinant of a sum of matrices (see appendix IX. 1 in \cite{Korepinbook}):
\begin{equation} \label{det_sum}
    \det (A+B)=\sum_{P_L,P_C}(-1)^{[P_L]+[P_C]}\det A_{P_L P_C}\det B_{P_L P_C}.
\end{equation}
Here the sum is over all partitions $P_L,P_C$ of the rows and columns into partitions of type $A$ and $B$. There are $n_A$ elements in partition $A$ and $n_B$ elements in partition $B,$ and $n_A+n_B=\mathrm{dim}(A)=\mathrm{dim}(B)$. The $A$ partition of the rows need not contain the same elements as the $A$ partition of the columns. The determinant $\det A_{P_L P_C}$ is the determinant of the matrix $A$ where all the rows and columns corresponding to the partitions of type $B$ have been removed. Finally, $(-1)^{[P_L]}$ is the sign of the permutation that puts all the rows corresponding to the elements in the partition of type $A$ first in the full $A$ matrix, while $(-1)^{[P_C]}$ is the sign of the permutation that puts all the elements corresponding to the partition of type $A$ first.

One can show that the sign $(-1)^P$ in \eqref{singlet_det1} is the same as the sign that is required to make use of the Laplace formula \eqref{det_sum}. The extra sign $(-1)^{\frac{n_IL}{2}}$ can be absorbed into the determinants, as can all vacuum eigenvalue factors, so the only problematic parts are the factors
\begin{equation}
    f(\lambda^{II}_k,\lambda^I_j)\frac{\lambda_j^{I}-\lambda_{k}^{II}}{\lambda_j^{I}+\lambda_{k}^{II}}.
\end{equation}
They can be dealt with by introducing a pair of auxiliary quantum fields $q,p$ with the only non-trivial commutation relations
\begin{equation}
    [p(\lambda),q(\mu)]=\ln\left(f(\mu,\lambda)\frac{\lambda-\mu}{\lambda+\mu}\right)
\end{equation}
and a vacuum $|0)$ satisfying
\begin{equation}
    p(\lambda)|0)=0,\quad (0|q(\lambda)=0.
\end{equation}
This kind of manipulation is done in \cite{Korepinbook} for off-shell overlaps between Bethe states.
Using the auxiliary quantum fields the overlap may be written as
\begin{equation}
     \bra{C_S}\prod_{n=1}^{\frac{L}{2}}B(\lambda_n)\ket{0}=\frac{\prod_{k=1}^{\frac{L}{2}} a_2(\lambda_k)d_2(\lambda_k)}{\prod_{1\leq k <l\leq \frac{L}{2}}(\nu_l-\nu_k)(\lambda_l-\lambda_k)}(0|\det S|0)
\end{equation}
where the matrix $S$ is given by
\begin{align}
    S_{k,l}=-a_2(\lambda_k)\frac{a_1(\lambda_k)d_1(\lambda_k)}{a_2(\lambda_k)d_2(\lambda_k)}&\frac{1}{(-\lambda_k+\nu_l-\frac{i}{2})(-\lambda_k+\nu_l+\frac{i}{2})} e^{p(\lambda_k)}\nonumber\\&+d_1(\lambda_k)\frac{1}{(\lambda_k+\nu_l-\frac{i}{2})(\lambda_k+\nu_l+\frac{i}{2})} e^{q(\lambda_k)}.
\end{align}
This expression can be simplified further by factoring out all factors $e^{p(\lambda_k)}$. 
This leads to a new dual vacuum
\begin{equation}
    (\Tilde{0}|=(0|\exp{\sum_{k=1}^{\frac{L}{2}} p(\lambda_k)}
\end{equation}
satisfying 
\begin{equation}
    (\Tilde{0}|q(\mu)=(\tilde{0}|\sum^{\frac{L}{2}}_{k=1}[p(\lambda_k),q(\mu)]=(\tilde{0}|\sum^{\frac{L}{2}}_{k=1}\ln\left(f(\mu,\lambda_k)\frac{\lambda_k-\mu}{\lambda_k+\mu}\right).
\end{equation}
Defining new fields $p',q',\varphi$ such that
\begin{equation}
    p'(\lambda)=p(\lambda),\quad q'(\mu)=q(\mu)-\sum^{\frac{L}{2}}_{k=1}\ln\left(f(\mu,\lambda_k)\frac{\lambda_k-\mu}{\lambda_k+\mu}\right),\quad \varphi(\lambda) = q'(\lambda)-p'(\lambda)
\end{equation}
where $\varphi$ has commutation relation
\begin{equation}
    [\varphi(\lambda),\varphi(\mu)]=-\ln\left(f(\lambda,\mu)f(\mu,\lambda)\frac{1}{(\lambda+\mu)^2}\right)
\end{equation}
the overlap becomes
\begin{equation}
    \bra{C_S}\prod_{n=1}^{\frac{L}{2}}B(\lambda_n)\ket{0}=\frac{\prod_{k=1}^{\frac{L}{2}} a_2(\lambda_k)d_2(\lambda_k)}{\prod_{1\leq k <l\leq \frac{L}{2}}(\nu_l-\nu_k)(\lambda_l-\lambda_k)}(\Tilde{0}|\det \Tilde{S}|0)
\end{equation}
with the matrix $\tilde{S}$ given by
\begin{align}
    \tilde{S}_{k,l}=-a_2(\lambda_k)&\frac{a_1(\lambda_k)d_1(\lambda_k)}{a_2(\lambda_k)d_2(\lambda_k)}\frac{1}{(-\lambda_k+\nu_l-\frac{i}{2})(-\lambda_k+\nu_l+\frac{i}{2})} \nonumber\\&+d_1(\lambda_k)\frac{1}{(\lambda_k+\nu_l-\frac{i}{2})(\lambda_k+\nu_l+\frac{i}{2})} e^{\varphi(\lambda_k)}\prod_{n\neq k}f(\lambda_k,\lambda_n)\frac{\lambda_n-\lambda_k}{\lambda_n+\lambda_k}.
\end{align}
This formula only involves one auxiliary quantum field $\varphi$.
\section{Conclusions}\label{conclusions}
We have discussed two classes of integrable crosscap states and derived off-shell formulae for their overlaps with Bethe states. 
The off-shell formulae we obtained may potentially be used to derive the on-shell overlaps. For example, it might be possible to use a procedure like the one that was used in \cite{Jiang2020} to perform the on-shell limit of VBS overlaps. The procedure amounts to taking the limit where the rapidities become parity invariant and deriving a recursion relation for the on-shell overlap. This was recently done \cite{gombor_crosscap} for the crosscap overlap in the more general $\mathfrak{gl}(n)$ case. The on-shell limit is expected to be more difficult for the crosscap singlet overlaps, because if $\frac{L}{2}$ is even they are non-zero only if the Bethe state is singular, as we discussed in section \ref{sect:on-shell_overlaps}. In addition, it is of course not know if the crosscap singlet overlaps take the expected form \eqref{gen_int_overlap}, and it would be interesting to see if it does. But even if no such on-shell formula is derived one can take the parity invariant limit and regularize the singularities before inserting the Bethe roots into the off-shell formulae. The on-shell formula would thus still be sums over partitions.

In the on-shell limits these formulae may potentially be of use in the study of quantum quenches, as was discussed in the introduction. In that case one would need to take the thermodynamic limit, which can be done for the crosscap overlap, since it has the standard overlap form \eqref{gen_int_overlap}. The thermodynamic limit of an overlap with this form was computed in \cite{Brockmann_2014}. 

Finally, integrable overlaps have been useful in the AdS/CFT correspondence, as we discussed in the introduction. It is not known if crosscap state overlaps appears in any defect configuration in $\mathcal{N}=4\ \mathrm{SYM},$ but work in this direction is in progress \cite{caetano_2022}.
\subsection*{Acknowledgement}
I am grateful to my former supervisor Kostya Zarembo for the many useful discussions throughout this project.
\subsection*{Note added}
After this work was completed \cite{gombor_crosscap} appeared where equation \eqref{cc_off-shell} is derived for $\mathfrak{gl}(n)$ spin chains in the same way as in the present article.
\printbibliography

@article{de-Leeuw_2015,
   title={One-point functions in defect CFT and integrability},
   volume={2015},
   ISSN={1029-8479},
   %url={http://dx.doi.org/10.1007/JHEP08(2015)098},
   DOI={10.1007/jhep08(2015)098},
   number={8},
   journal={Journal of High Energy Physics},
   publisher={Springer Science and Business Media LLC},
   author={de Leeuw, Marius and Kristjansen, Charlotte and Zarembo, Konstantin},
   eprint = {1506.06958},
  primaryClass = {hep-th},
  archivePrefix = {arXiv},
   year={2015}
}

@article{gombor_2021,
      title={On exact overlaps for $\mathfrak{gl}(N)$ symmetric spin chains}, 
      author={Tamás Gombor},
      year={2021},
      eprint={2110.07960},
      archivePrefix={arXiv},
      primaryClass={hep-th}
}

@article{Foda_2016,
   title={Overlaps of partial Néel states and Bethe states},
   volume={2016},
   ISSN={1742-5468},
   %url={http://dx.doi.org/10.1088/1742-5468/2016/02/023107},
   DOI={10.1088/1742-5468/2016/02/023107},
   number={2},
   journal={Journal of Statistical Mechanics: Theory and Experiment},
   publisher={IOP Publishing},
   author={Foda, O and Zarembo, K},
   year={2016},
   pages={023107},
   eprint={1512.02533},
    archivePrefix={arXiv},
    primaryClass={hep-th}
}

@article{caetano_2022,
	doi = {10.1007/s10955-022-02914-6},
  
	%url = {https://doi.org/10.1007%2Fs10955-022-02914-6},
  
	year = 2022,
  
	publisher = {Springer Science and Business Media {LLC}
},
  
	volume = {187},
  
	number = {3},
  
	author = {Jo{\~{a}}o Caetano and Shota Komatsu},
  
	title = {Crosscap States in Integrable Field Theories and Spin Chains},
  
	journal = {Journal of Statistical Physics},
	eprint={2111.09901},
      archivePrefix={arXiv},
      primaryClass={hep-th}
}

@article{kristjansen_2021,
      title={Duality Relations for Overlaps of Integrable Boundary States in AdS/dCFT}, 
      author={Charlotte Kristjansen and Dennis Müller and Konstantin Zarembo},
      year={2021},
      eprint={2106.08116},
      archivePrefix={arXiv},
      primaryClass={hep-th}
}

@article{deLeeuw2020,
   title={One-point functions in AdS/dCFT},
   volume={53},
   ISSN={1751-8121},
   %url={http://dx.doi.org/10.1088/1751-8121/ab15fb},
   DOI={10.1088/1751-8121/ab15fb},
   number={28},
   eprint = {1908.03444},
  primaryClass = {hep-th},
  archivePrefix = {arXiv},
   journal={Journal of Physics A: Mathematical and Theoretical},
   publisher={IOP Publishing},
   author={de Leeuw, Marius},
   year={2020},
   pages={283001}
   }

@article{Gombor_2021/03,
   title={On factorized overlaps: Algebraic Bethe Ansatz, twists, and separation of variables},
   volume={967},
   ISSN={0550-3213},
   %url={http://dx.doi.org/10.1016/j.nuclphysb.2021.115390},
   DOI={10.1016/j.nuclphysb.2021.115390},
   journal={Nuclear Physics B},
   publisher={Elsevier BV},
   author={Gombor, Tamás and Pozsgay, Balázs},
   year={2021},
   pages={115390},
   eprint={2101.10354},
      archivePrefix={arXiv},
      primaryClass={cond-mat}}

@article{Jiang2020,
   title={On exact overlaps in integrable spin chains},
   volume={2020},
   ISSN={1029-8479},
   %url={http://dx.doi.org/10.1007/JHEP06(2020)022},
   DOI={10.1007/jhep06(2020)022},
   number={6},
   journal={Journal of High Energy Physics},
   publisher={Springer Science and Business Media LLC},
   author={Jiang, Yunfeng and Pozsgay, Balázs},
   year={2020},
   eprint={2002.12065},
      archivePrefix={arXiv},
      primaryClass={cond-mat}}

@book{Korepinbook,
publisher = {Cambridge University Press},
isbn = {9780521586467},
year = {2010},
title = {Quantum inverse scattering method and correlation functions },
language = {eng},
author = {Korepin, V. E and Izergin, A. G and Bogoliubov, N. M},
}

@article{Rahul1998,
  %doi = {10.48550/ARXIV.COND-MAT/9804210},
  
  %url = {https://arxiv.org/abs/cond-mat/9804210},
  
  author = {Siddharthan, Rahul},
  
  keywords = {Strongly Correlated Electrons (cond-mat.str-el), Exactly Solvable and Integrable Systems (nlin.SI), FOS: Physical sciences, FOS: Physical sciences},
  
  title = {Singularities in the Bethe solution of the XXX and XXZ Heisenberg spin chains},
  
  %publisher = {arXiv},
  
  year = {1998},
  eprint={9804210},
  archivePrefix={arXiv},
  primaryClass={cond-mat},
  copyright = {Assumed arXiv.org perpetual, non-exclusive license to distribute this article for submissions made before January 2004}
}

@article{Nepomechie_2013,
	doi = {10.1088/1751-8113/46/32/325002},
	%url = {https://doi.org/10.1088/1751-8113/46/32/325002},
	year = 2013,
	publisher = {{IOP} Publishing},
	volume = {46},
	number = {32},
	pages = {325002},
	eprint={1304.7978},
	archivePrefix={arXiv},
	primaryClass={het-th},
	author = {Rafael I Nepomechie and Chunguang Wang},
	title = {Algebraic Bethe ansatz for singular solutions},
	journal = {Journal of Physics A: Mathematical and Theoretical},
	abstract = {The Bethe equations for the isotropic periodic spin-1/2 Heisenberg chain with N sites have solutions containing ±i/2 that are singular: both the corresponding energy and the algebraic Bethe ansatz vector are divergent. Such solutions must be carefully regularized. We consider a regularization involving a parameter that can be determined using a generalization of the Bethe equations. These generalized Bethe equations provide a practical way of determining which singular solutions correspond to eigenvectors of the model.}
}

@article{Faddevnotes,
  %doi = {10.48550/ARXIV.HEP-TH/9605187},
  
  eprint = {hep-th/9605187},
  archivePrefix = {arXiv},
  author = {Faddeev, L. D.},
  
  keywords = {High Energy Physics - Theory (hep-th), FOS: Physical sciences, FOS: Physical sciences},
  
  title = {How Algebraic Bethe Ansatz works for integrable model},
  
  booktitle = {Les
Houches School of Physics: Astrophysical Sources of Gravitational Radiation},
  
  publisher = {arXiv},
  
  year = {1996},
  
  copyright = {Assumed arXiv.org perpetual, non-exclusive license to distribute this article for submissions made before January 2004}
}

@article{Slavnovnotes,
  %doi = {10.48550/ARXIV.1804.07350},
  
  eprint = {1804.07350},
  primaryClass = {math-ph},
  archivePrefix = {arXiv},
  author = {Slavnov, N. A.},
  
  keywords = {Mathematical Physics (math-ph), FOS: Physical sciences, FOS: Physical sciences},
  
  title = {Algebraic Bethe ansatz},
  
  publisher = {arXiv},
  
  year = {2018},
  
  copyright = {arXiv.org perpetual, non-exclusive license}
}

@article{Korepin_1982,
issn = {0010-3616},
journal = {Communications in mathematical physics},
pages = {391--418},
volume = {86},
publisher = {Springer-Verlag},
number = {3},
year = {1982},
title = {Calculation of norms of Bethe wave functions},
copyright = {Copyright 1982 Springer-Verlag},
language = {eng},
author = {Korepin, V. E.},
doi = {10.1007/BF01212176}
}

@article{Piroli_2017,
title = {What is an integrable quench?},
journal = {Nuclear Physics B},
volume = {925},
pages = {362-402},
year = {2017},
issn = {0550-3213},
eprint = {1709.04796},
primaryClass = {cond-mat},
archivePrefix = {arXiv},
doi = {10.1016/j.nuclphysb.2017.10.012},
%url = {https://www.sciencedirect.com/science/article/pii/S0550321317303413},
author = {Lorenzo Piroli and Balázs Pozsgay and Eric Vernier},
abstract = {Inspired by classical results in integrable boundary quantum field theory, we propose a definition of integrable initial states for quantum quenches in lattice models. They are defined as the states which are annihilated by all local conserved charges that are odd under space reflection. We show that this class includes the states which can be related to integrable boundary conditions in an appropriate rotated channel, in loose analogy with the picture in quantum field theory. Furthermore, we provide an efficient method to test integrability of given initial states. We revisit the recent literature of global quenches in several models and show that, in all of the cases where closed-form analytical results could be obtained, the initial state is integrable according to our definition. In the prototypical example of the XXZ spin-s chains we show that integrable states include two-site product states but also larger families of matrix product states with arbitrary bond dimension. We argue that our results could be practically useful for the study of quantum quenches in generic integrable models.}
}

@article{Pozsgay_2018,
	doi = {10.1088/1742-5468/aabbe1},
  
  
	year = 2018,
  
	publisher = {{IOP} Publishing},
  
	volume = {2018},
  
	number = {5},
  eprint = {1801.03838},
  primaryClass = {cond-mat},
  archivePrefix = {arXiv},
	pages = {053103},
  
	author = {Pozsgay, B.},
  
	title = {Overlaps with arbitrary two-site states in the {XXZ} spin chain},
  
	journal = {Journal of Statistical Mechanics: Theory and Experiment}
}

@article{Brockmann_2014,
	doi = {10.1088/1751-8113/47/14/145003},
  
	%url = {https://doi.org/10.1088%2F1751-8113%2F47%2F14%2F145003},
  
	year = 2014,
  
	publisher = {{IOP} Publishing},
  
	volume = {47},
  
	number = {14},
  
	pages = {145003},
  
  	eprint = {1401.2877},
  primaryClass = {cond-mat},
  archivePrefix = {arXiv},
  
	author = {M Brockmann and J De Nardis and B Wouters and J-S Caux},
  
	title = {A Gaudin-like determinant for overlaps of N{\'{e}
}el and {XXZ} Bethe states},
  
	journal = {Journal of Physics A: Mathematical and Theoretical}
}

@article{Heisenberg_1928,
    year={1928},
    title={Zur Theorie des Ferromagnetismus},
    journal={Zeitschrift für Physik},
    pages={619-636},
    volume={49},
    number={9},
    doi = {10.1007/BF01328601},
    author={Heisenberg, W.}
}

@article{Eisert_2015,
	doi = {10.1038/nphys3215},
  
	%url = {https://doi.org/10.1038%2Fnphys3215},
  
	year = 2015,
  
	publisher = {Springer Science and Business Media {LLC}
},
  
	volume = {11},
  
	number = {2},
  
	pages = {124--130},
  
	author = {J. Eisert and M. Friesdorf and C. Gogolin},
  
	title = {Quantum many-body systems out of equilibrium},
	
	eprint = {1408.5148v2},
  primaryClass = {quant-ph},
  archivePrefix = {arXiv},
  
	journal = {Nature Physics}
}

@article{Gogolin_2016,
	doi = {10.1088/0034-4885/79/5/056001},
  
	%url = {https://doi.org/10.1088\%2F0034-4885\%2F79\%2F5\%2F056001},
  
	year = 2016,
  
	publisher = {{IOP} Publishing},
  
	volume = {79},
  
	number = {5},
  
	pages = {056001},
	
	  	eprint = {1503.07538},
  primaryClass = {quant-ph},
  archivePrefix = {arXiv},
  
	author = {Christian Gogolin and Jens Eisert},
  
	title = {Equilibration, thermalisation, and the emergence of statistical mechanics in closed quantum systems},
  
	journal = {Reports on Progress in Physics}
}

@article{Calabrese_2006,
	doi = {10.1103/physrevlett.96.136801},
  
	%url = {https://doi.org/10.1103\%2Fphysrevlett.96.136801},
  
	year = 2006,
  
	publisher = {American Physical Society ({APS})},
  
	volume = {96},
  
	number = {13},
  
	author = {Pasquale Calabrese and John Cardy},
	
    eprint = {cond-mat/0601225},
  archivePrefix = {arXiv},
  
	title = {Time Dependence of Correlation Functions Following a Quantum Quench},
  
	journal = {Physical Review Letters}
}

@article{Pozsgay_2014,
	doi = {10.1088/1742-5468/2014/06/p06011},
  
	%url = {https://doi.org/10.1088\%2F1742-5468\%2F2014\%2F06\%2Fp06011},
  
	year = 2014,
  
	publisher = {{IOP} Publishing},
  
	volume = {2014},
  
	number = {6},
  
	pages = {P06011},
  
  eprint = {1309.4593},
  primaryClass = {cond-mat},
  archivePrefix = {arXiv},
  
	author = {Bal{\'{a}
}zs Pozsgay},
  
	title = {Overlaps between eigenstates of the {XXZ} spin-1/2 chain and a class of simple product states},
  
	journal = {Journal of Statistical Mechanics: Theory and Experiment}
}

@article{Wouters_2014,
	doi = {10.1103/physrevlett.113.117202},
  
	%url = {https://doi.org/10.1103\%2Fphysrevlett.113.117202},
  
	year = 2014,
  
	publisher = {American Physical Society ({APS})},
  
	volume = {113},
  
	number = {11},
	
	  eprint = {1405.0172},
  primaryClass = {cond-mat},
  archivePrefix = {arXiv},
  
	author = {B. Wouters and J. De Nardis and M. Brockmann and D. Fioretto and M. Rigol and J.-S. Caux},
  
	title = {Quenching the Anisotropic Heisenberg Chain: Exact Solution and Generalized Gibbs Ensemble Predictions},
  
	journal = {Physical Review Letters}
}

@article{Buhl_Mortensen_2016,
	doi = {10.1007/jhep02(2016)052},
  
	%url = {https://doi.org/10.1007\%2Fjhep02\%282016\%29052},
  
	year = 2016,
  
	publisher = {Springer Science and Business Media {LLC}
},
  
	volume = {2016},
  
	number = {2},
	
	eprint = {1512.02532},
  primaryClass = {het-th},
  archivePrefix = {arXiv},
  
	author = {Isak Buhl-Mortensen and Marius de Leeuw and Charlotte Kristjansen and Konstantin Zarembo},
  
	title = {One-point functions in {AdS}/{dCFT} from matrix product states},
  
	journal = {Journal of High Energy Physics}
}

@article{Kristjansen_2020,
author={Kristjansen, C. and Müller, D. and Zarembo, K.},
title={Integrable boundary states in D3-D5 dCFT: beyond scalars},
journal={Journal of High Energy Physics},
year={2020},
eprint = {2005.01392},
  primaryClass = {het-th},
  archivePrefix = {arXiv},
issue={103},
doi = {10.1007/JHEP08(2020)103}
%url={https://doi.org/10.1007/JHEP08(2020)103}
}

@article{Minahan_2003,
	doi = {10.1088/1126-6708/2003/03/013},
  
	%url = {https://doi.org/10.1088/1126-6708/2003/03/013},
  
	year = 2003,
  
	publisher = {Springer Science and Business Media {LLC}
},
  
	volume = {2003},
  
	number = {03},
	pages = {013},
	
  eprint = {0212208},
  primaryClass = {het-th},
  archivePrefix = {arXiv},
  
	author = {Joseph A Minahan and Konstantin Zarembo},
  
	title = {The Bethe-ansatz for Script N = 4 super Yang-Mills},
  
	journal = {Journal of High Energy Physics}
}

@article{avdeev,
    author = {L.V. Avdeev and A.A. Vladimirov},
    title = {Exceptional solutions to the Bethe ansatz equations},
    journal = {Theor Math Phys},
    pages = {1071-1079},
    year = {1986},
    volume = {69},
    doi = {https://doi.org/10.1007/BF01037864}
}

@article{gombor_crosscap,
  %doi = {10.48550/ARXIV.2207.10598},
  
  %url = {https://arxiv.org/abs/2207.10598},
  
  author = {Gombor, Tamas},
  
  eprint = {2207.10598},
  primaryClass = {het-th},
  archivePrefix = {arXiv},
  
  keywords = {High Energy Physics - Theory (hep-th), Mathematical Physics (math-ph), FOS: Physical sciences, FOS: Physical sciences},
  
  title = {Integrable crosscap states in $\mathfrak{gl}(N)$ spin chains},
  
  publisher = {arXiv},
  
  year = {2022},
  
  copyright = {arXiv.org perpetual, non-exclusive license}
}
\end{document}